\documentclass[12pt]{article}
\usepackage{amsfonts,amsmath} 

\date{ \vspace*{-.1truein} {\small  July 12, 2000; revised Sept. 12,
2000. }}

\input epsf     
\def\abst#1{\begin{minipage}{5.5in}
{\noindent   \normalsize
{\bf Abstract} #1}  
\end{minipage}  }

\def\be{\begin{equation}}
\def\ee{\end{equation}}
\def\bea{\begin{eqnarray}}
\def\eea{\end{eqnarray}}

\newcommand{\eq}[1]{eq.~(\ref{#1})}

\def\var{{\rm Var}}

\def\blackbox{{\vrule height 1.3ex width 1.0ex depth -.2ex}
       \hskip 1.5truecm}

\def\E{{\mathbb E}}    

\def\P{\operatorname{Prob}} 
\def\R{{\mathbb R}}    

\def\Z{{\mathbb Z}}    

\newcounter{masectionnumber}
\newcommand{\masect}[1]{\setcounter{equation}{0}
\refstepcounter{masectionnumber} \vspace{1truecm plus 1cm} \noindent
    {\large\bf \arabic{masectionnumber}. #1}\par \vspace{.2cm}
      \addcontentsline{toc}{section}{\arabic{masectionnumber}. #1}
    }

\newcounter{masubsectionnumber}[masectionnumber]
\newcommand{\masubsect}[1]{
    \refstepcounter{masubsectionnumber} \vspace{.5cm} \noindent
    {\large\em \arabic{masectionnumber}.\alph{masubsectionnumber} #1}
    \par\vspace*{.2truecm}
    \addcontentsline{toc}{subsection}
    {\arabic{masectionnumber}.\alph{masubsectionnumber}\hspace{.1cm}
        #1}
    }

\newtheorem{lem}{Lemma}[masectionnumber]
\newtheorem{thm}[lem]{Theorem}

\newtheorem{df}{Definition}[masectionnumber]

\newenvironment{proof_of}[1]{
    \noindent{\bf Proof of #1:} \hspace*{1em} }{
    \hfill \blackbox\bigskip}

    \newcounter{masubapp}[masectionnumber]
\begin{document}
\title {  \vspace*{-1in}
{\Large Bounded Fluctuations and Translation Symmetry Breaking} \\
{\Large in One-Dimensional Particle Systems}{$^*$}
}

\author{M. Aizenman $^{a}$ \ \  \ \  S. Goldstein $^b$ \ \  \ \
   J. L. Lebowitz $^{b}$     \vspace*{0.25truein} \\
\normalsize \it  $^{(a)}$ Departments of Physics and Mathematics,
Princeton University, \vspace*{-.1truein} \\
\normalsize \it Jadwin Hall, P. O. Box 708, Princeton, NJ 08544. \\
\normalsize \it  $^{(b)}$ Departments of Mathematics and
Physics, Rutgers University,  \vspace*{-.1truein} \\
\normalsize \it    110 Frelinghuysen Road, Piscataway,
NJ 08854-8019.  }  
\maketitle
\thispagestyle{empty}        

 \begin{abstract} \abst{ We present general results for one-dimensional
systems of point charges (signed point measures)
on the line with a translation invariant
distribution $\mu$ for which the variance of the total charge in an
interval is uniformly bounded (instead of increasing with the interval
length).  When the charges are restricted to multiples of a common unit,
and their average charge density does not vanish, then the boundedness of
the variance implies translation-symmetry breaking --- in the sense that
there exists a function of the charge configuration that is nontrivially
periodic under translations --- and hence that $\mu$ is not ``mixing.''
Analogous results are formulated also for one dimensional lattice systems
under some constraints on the values of the charges at the lattice sites
and their averages.  The general results apply to one-dimensional Coulomb
systems, and to certain spin chains, putting on common grounds different
instances of symmetry breaking encountered there.}

 \end{abstract}

\noindent $^*$Dedicated to the memory of J. M. (Quin) Luttinger, a master
of one-dimensional systems and much more.

%

\masect{Introduction}
\label{sect:intro}
%

To fluctuate is normal, and normally fluctuations grow like the square root
of the volume.  There are however non-trivial exceptions to this rule;
among the notable examples are systems of charges with Coulomb interaction
and certain quantum spin chains in their ground states.  Curiously, the
known one-dimensional systems with uniformly bounded fluctuations in the
total charge or component of spin in an interval exhibit also other
exceptional properties, such as translation symmetry breaking (in jellium
and quantum spin chains), or non-uniqueness of the Gibbs state (in the
two-component neutral Coulomb system).  We prove here that this is not
accidental: under some general assumptions, bounded fluctuations in a
one-dimensional particle system imply the existence there of a periodic, or
in the lattice case a quasi-periodic, structure.  In particular, in such
systems the correlations of certain local variables do not decay.

We briefly describe the examples mentioned above to which our results
apply.

\noindent{\bf a) Coulomb systems:} In general terms, a Coulomb system
consists of several ($m$) species of particles with charges $q_\alpha$,
$\alpha = 1,...,m$.  For a classical system containing $N_\alpha$ particles
of species  $\alpha$
 in a finite domain $V\subset {\Bbb R}^d$, the
(configurational) Gibbs
canonical distribution is the probability measure on the space of
configurations $\{(x_j, \alpha_j)\}_{j=1,\ldots,N},$\break $ \ x_j \in V,\
\alpha_j=1,\ldots,m$, and $N=\sum_{\alpha=1}^m N_{\alpha}$,
given by the density $\exp[-\beta U(\{(x_j, \alpha_j)\}) ] / Z$, with
\begin{equation}
U(\{(x_j, \alpha_j)\}) \ =\
\sum_{i\ne j} q_j q_i \ V_{C}(x_i - x_j) +  \sum_{j} q_j V_{bg}(x_j)
+ \sum_{i,j}V_{sr}^{(\alpha_i, \alpha_j)}(x_i - x_j)  \; ,
\end{equation}
where $q_j \equiv q_{\alpha_j}$, $V_{C}(x)$ is the Coulomb
potential,
satisfying $\Delta V_{C}(x) = -\delta(x)$, $V_{bg}(x) $ is the
potential induced by a (uniform)
 background charge, of {\em charge} density
$\rho_{bg}$, and $V_{sr}^{(\alpha,\alpha')}(x) $ is a short range
interaction between particles of
species $\alpha$ and $\alpha'$.  For quantum systems
the measure on the space of configurations is given by a more complicated
formula; for details, we refer the reader to the recent review article by
Brydges and Martin~\cite{BM}.

Under suitable conditions, such measures,
with $\{ q_j \}$ and $\rho_{bg}$ not all of the same sign,
 admit translation invariant infinite volume limits.  In the
limiting states  the total
charge density, including the background, is zero, i.e., the
{\em particle densities}  $\nu_{\alpha}$
satisfy $\sum_{\alpha} q_{\alpha} \nu_{\alpha} + \rho_{bg} = 0$
[1-3].
It is further expected, and proven in some cases, including all classical
one-dimensional Coulomb systems \cite{Lenard,AM}, that with respect to
these limiting measures the variance of the net charge in a region $\Lambda
$ increases with $\Lambda $ only like its surface area~\cite{vB,L}.  That is,
\begin{equation}\label{lambda}
\var(Q_\Lambda) :=\langle
Q_\Lambda^2 \rangle - \langle Q_\Lambda \rangle^2
\ \sim \ |\partial \Lambda| \qquad
\mbox{(as $\Lambda \nearrow {\Bbb R}^d $) },
\end{equation}
where $Q_\Lambda = \sum_{x_i  \in \Lambda} q_{\alpha_i}$.
In one dimension this corresponds to the statement that the variance of the
charge in an interval $I$ remains bounded as $|I| \to \infty$.
Analogous statements apply to the case where the charges are
restricted to lattice sites~\cite{Gaudin}. (Eq. (\ref{lambda}) follows from
the ''zero sum rule,'' which holds whenever the charge correlations have
sufficiently rapid decay \cite{BM,vB}).

A simple example of a Coulomb system is the so-called {\em jellium} model,
 or the {\em one-component plasma} (OCP), with particles of unit charge
 dispersed in a negatively charged uniform background.  For this system
 surface growth of charge fluctuations, which now correspond to
 particle-number fluctuations, has been established
  in $d = 1$ for all temperatures \cite{Lenard,AM}
 and in $d \geq 2$ at high temperatures $(\beta << 1)$.
 \footnote{
The statement for the high temperatures  follows
through the combination of the results of \cite{Imbrie} and \cite{vB}.
In $d=2$, surface growth can also be verified
for the exactly solvable OCP at
$\beta =2$ \cite{Janco}; numerical calculations support
it for all $\beta$ \cite{L}.}

For the one-dimensional jellium (with the $1$D Coulomb potential
$V_{C}(x) = -\frac12|x|$) it is also known that the limits of the  Gibbs
measures exhibit ``translation-symmetry
breaking''~\cite{Kunz,BL,AM,Lug-Mart}.
The periodic structure found there
has been regarded as an example of the `Wigner lattice.'
However, in  this case the symmetry breaking
was also understood to be related
to the boundedness of the charge  fluctuations,
the two being connected through the properties of the electric
field~\cite{AM} (for which the existence of the limit
was previously established in ref.~\cite{Lenard}).
Here we show that this relation is an example of a more general
phenomenon.

\noindent {\bf b) Spin chains:}
Another example of a one-dimensional system with reduced
fluctuations  is provided by the ground state of certain
quantum spin systems.  These consist of `chains'
(also of interest are arrangements into `ladders') of
 quantum spins
$\{ \underline{\sigma}_{n} \}_{n \in \Z}$,
of a common spin $S$, such that $2 S$ is an integer,
with the Hamiltonian
\begin{equation}
H \ = \  -  \sum_{n\in \Z}  P^{(0)}_{n,n+1} \; ,
\end{equation}
where $ P^{(0)}_{n,n+1} $ is the projection onto the singlet state, i.e.,
on the subspace in which
$(\underline{\sigma}_{n} + \underline{\sigma}_{n+1})^2 =0$.
This class of Hamiltonians was introduced by
Affleck~\cite{Affleck} as an interesting extension of the spin $1/2$
Heisenberg antiferromagnetic spin chain. (It is  also  related to
the classical Potts antiferromagnet with $Q= 2S+1$ \cite{AN}).

The `$z$-components' of the spins, $\{ \sigma^{(3)}_n \}$, form a family of
commuting observables.  Thus in any quantum state their joint distribution
may be described by a `classical' probability measure on the space of
configurations of one-component spin variables with values ranging over
$\{-S,\ -S+1, \ldots,\ S\}$.  To distinguish these from the full quantum spin
variables let us refer to their values as `charges.'  For these
systems the ground state ($|0\rangle$) admits a
representation in which the charges are organized into
\underline{neutral clusters}~\cite{AN}, i.e., randomly organized
clusters of sites for which $\sum \sigma^{(3)}
=0$, with relative spin flip symmetry between different clusters.
The clusters can intermingle, so this perspective can be of value
only when the distribution of these clusters is such that for any
interval $I$, the number of sites in $I$ belonging to
clusters  which are not entirely contained in the given interval
is seldom large -- in a sense uniform in $I$.
A condition implying that this scenario is realized
is rapid decay of correlations:
\begin{equation}
\sum_{n>0} n \,
| \langle0|\  \sigma^{(3)}_0 \sigma^{(3)}_{n} \ |0\rangle |  \
<\ \infty  \; .
\label{eq:dimerization}
\end{equation}
One can show, using the cluster representation, that when
\eq{eq:dimerization} holds
the two point function satisfies the neutrality condition
$\sum_{x} \langle0|\  \sigma^{(3)}_0 \sigma^{(3)}_{x} \ |0\rangle =
0$, which is analogous to the ``zero sum rule'' of Coulomb
systems~\cite{BM,vB,L},
and the block spins
$S_{I}= \sum_{n\in I} {\sigma}_{n}^{(3)}$ have  uniformly bounded
variance, with
\begin{equation}
\langle0|\  |S_{I}|^2 \ |0\rangle| \ \le \ 2 \sum_{n>0} n \,
| \langle0|\  \sigma^{(3)}_0 \sigma^{(3)}_{n} \ |0\rangle| \  \; .
\end{equation}
The validity of the condition (\ref{eq:dimerization}) depends on $S$; exact
calculations indicate that (\ref{eq:dimerization}) is satisfied for all $S
\ge 1$, but not for $S=1/2$~\cite{Affleck,Bat-Bar,Kl}.

In \cite{AN} it was shown, using the aforementioned cluster
representation, that
for ``half-integer'' spins (with $2S$ an odd integer greater than one)
condition (\ref{eq:dimerization}) implies, in addition to the bound
on block spin fluctuations, also that the translation invariant
infinite volume ground state decomposes into a mixture
of two states of period $2$.  The symmetry breaking which occurs in the
above spin systems is akin to {\em dimerization}, though the
neutral clusters need not consist just of pairs of neighboring spins.
(A notable fact is that the lack of mixing in the  ground state is
expressed through other correlators than the two-point
correlation function, since the spin-spin correlation
does decay to zero.)

The general
results presented here place this translation symmetry breaking
within a broader context, as we show that it is
not coincidental that the same condition  (\ref{eq:dimerization})
implies, for  the quantum spin chains
 with {\em odd} values of $2S$,
both bounded variance of
the block spins and translation symmetry breaking.

\masect{Formulation and results}
\label{sect:statement}

Our main results relate reduced fluctuations to
{\em translation symmetry breaking}.  A more precise formulation
of the conclusion is that the state, given by a translation
invariant measure $\mu$ on the space of configurations, has a
{\em cyclic factor}.  Let us first introduce the relevant terminology.

\masubsect{Cyclic factors and symmetry breaking}

\begin{df}\label{def}
A system described by a probability measure  $\mu(d\omega)$,
on a space $\Omega$ [here, the space of charge configurations],
which is invariant under either the continuous
group of translations ( shifts $T_x$ by $x\in \R$)
or the group of lattice shifts ( $x\in \Z$) is said to have
a \underline{cyclic factor}, of period
$0< \lambda < \infty$, if
there is a  measurable function  $\phi(\omega)$
with values in $[0,2 \pi)$ that evolves  under the shifts by
\begin{equation}
\phi(T_x \omega) \ = \ \phi(\omega) - 2 \pi \, x /  \lambda
   \quad \mbox{({\rm mod} $2 \pi $), for a.e. $\omega$},
\label{eq:phi}
\end{equation}
with, in the lattice case, $\lambda^{-1}\notin\Z$.
\end{df}

An equivalent formulation
is that the  measure $\mu$ can be decomposed into a  mixture:
\begin{equation}\label{eq:cc}
\mu(\cdot) =  \int_0^{ 2 \pi} \frac{d \theta}{2 \pi} \ \  \mu_{\theta}(\cdot)
\end{equation}
of mutually singular measures which are cycled under the shifts $T_x$:
\begin{equation}\label{eq:mu}
T_x \ \mu_{\theta}  \  =
\mu_{(\theta - 2\pi x / \lambda)_{({\rm mod}\ 2\pi)} } \; , \qquad
d\theta{\rm -a.s.} .
\end{equation}
In the lattice case, if $\lambda$ is rational then $\frac{d\theta}{2 \pi}$
 in (\ref{eq:cc}) [and in (\ref{eq:mu})] should be replaced by, or
 understood as, a probability measure $\nu(d\theta)$ invariant under
 $\theta\longmapsto \theta - \frac{2\pi}{\lambda}\ ({\rm mod}\ 2\pi)$.

The two formulations are related by the observation that
in the decomposition (\ref{eq:cc})  the \underline{cyclic component}
$\mu_{\theta}(\cdot)$ is supported
by configurations $\omega$ with $\phi(\omega)=\theta$.

If the invariance is under the continuous group of translations, the
existence of a cyclic factor implies that the measure
$\mu$ is not ergodic under the smaller group of shifts by $\lambda$.
For lattice systems, which are invariant only under shifts by
multiples of the lattice period, if $\lambda$ is incommensurate with
the
period (here 1) then each of the lattice shifts may still act ergodically
on $\mu$ and the translation symmetry breaking
is  expressed only through the existence of a quasi-periodic
structure; if the two lengths are commensurate then there is loss of
ergodicity under shifts by some {\em multiple} of $\lambda$.

Nevertheless, in either the continuum or the lattice cases, the existence
of a cyclic factor implies that $\mu$ does not have good clustering
properties.  In particular, the {\em mixing} condition, which implies that
for any bounded function $g(\omega)$ with expectation $\int g\,d\mu =0$
\begin{equation}
 \int g\,T_x g\, d\mu \to 0 \; , \quad \mbox{as $|x|\to \infty$}\;,
\end{equation}
is not satisfied.

Let us note also that it
follows from (\ref{eq:phi}) that $\phi(\omega)$ is a tail function
(under translations), in fact
measurable with respect to the $\sigma$-algebra at $+\infty$ (or $-\infty$).
Thus, if
$\mu$ is a Gibbs state for some potential, or weakly
Gibbs~\cite{MS}, then
its cyclic components should also be Gibbs states, in the appropriate
sense, for that potential.

\masubsect{Results}

Our first result concerns systems of particles in the continuum. The two
cases of primary interest to us are: i) point processes, described by
measures on the space of locally finite particle configurations on the
line $\R$, and ii) point-charge processes on $\R$.  A process of either
type may be described in terms of a random atomic measure on $\R$ for which
the collection of atoms (the locations of the points or the charges) is
(a.s.) locally finite.  In the former case the measure, which will be
denoted by $N_{\omega}(I)$, is the number of particles in the
interval $I$; in the
latter case the measure is the sum of the point charges in $I$,
 $Q_{\omega}(I)=\sum_{x_i\in I}q_{\alpha_i}$, in the notation
 of example {\bf 1.a}.

\begin{thm}
Let $\mu(d\omega)$ be a translation invariant probability measure
describing a point process on $\R$.
If the variance of the number of points in an interval
$I$ (i.e., of the random variable $N_I(\omega)\equiv N_{\omega}(I)$)
is bounded uniformly in the size of the interval
\begin{equation}
\E\left([N_I - \E(N_I)]^2 \right) \ \le \ C \ (< \infty)
\label{eq:variance}
\end{equation}
then $\mu$ has a cyclic factor, of period
\begin{equation}
 \lambda \ = \ 1 /\, \E(N_{[0,1]}) \; .
\end{equation}

More generally, let $\mu$ describe a translation invariant point-charge
 process on the line, with the charge values  restricted to
multiples of a common unit $e$.
Assume that for some $\rho \ne 0$ the family of random
variables $\{ F_{\omega}(a,b) \}$, defined by
\begin{equation}
F_{\omega}(a,b) \ = \
Q_{\omega}((a,b]) - \rho (b-a)
\label{eq:F}
\end{equation}
where $-\infty < a < b < \infty$
and $Q_\omega(I)$ is the total charge in the interval $I$,  is
\underline{tight}.
Then the measure $\mu$  has a cyclic factor, of period
\begin{equation}
\lambda = e / \rho \; .
\end{equation}
\label{thm:1}
\end{thm}

The tightness of the family of random variables $\{ F_{\omega}(a,b)
\}_{a,b}$ means that the probabilities of large fluctuations have
asymptotically vanishing bounds that are uniform in $a,b \in \R$:
\begin{equation}
\P(\, | F_{\omega}(a,b)| \ge t \, ) \ \le \ p(t)
\end{equation}
with  $p(t) \searrow 0$ as $t \to \infty$.
A sufficient condition for this
is that the variances of $Q_I$ are uniformly bounded, as in
(\ref{eq:variance}).

The quantity $\rho$ introduced in Theorem~\ref{thm:1}
is the {\it asymptotic mean\/} of the charge density.
Under the tightness condition formulated above
it is well defined even if
$\E(\, |Q_{\omega}(I)|\, ) = \infty$.  Note also that ergodicity is
not assumed in Theorem~\ref{thm:1}.


Our second result deals with systems of charges $\{ q_k \}_{k \in \Z}$
on the one-dimensional lattice $\Z$.
Here, for any interval $I\subset \Z$, the charge in $I$ is given by
$ Q_{\omega}(I)\ = \ \sum_{k\in I} q_k  $, with $q_k  $ the charge
variable at the lattice site $k \in \Z$.

\begin{thm}
Let $\mu$ describe a translation invariant system of charges on the
lattice $\Z$ with the following properties:
\begin{enumerate}
\item[a.]
The charges ($q_k$) are restricted to values of the form
$ (\gamma + n \, e)  $
with $\gamma$ and $ e$ fixed,  and $n$ taking only integer values.
\item[b.]
For some $\rho \in \R $ the family of random
variables
$\{ F_{\omega}(a,b) \equiv Q_{\omega}((a,b]) - \rho (b-a) \}$,
with $ a, b\in \Z$ and $a<b$,
is tight.
 \item[c.]
Furthermore,
 \begin{equation}
  \rho - \gamma \ \ne \ 0 \ ({\rm mod} \ e) \; .
\label{eq:alpha}
 \end{equation}
\end{enumerate}
Then the measure $\mu$ has a cyclic factor, of period
\begin{equation}
\lambda = e / \alpha \; ,
\end{equation}
with $\alpha$ defined by
\begin{equation}
\alpha \in [0,e) \;, \qquad  \alpha \ = \ \rho - \gamma \ ({\rm
mod} \  e) \; .
\end{equation} \label{thm:2}
\end{thm}

Before we turn to the proof, let us note that the conditions
{\em a.\/}  and {\em c.\/}
 are satisfied in the following examples:
\begin{enumerate}
\item[Ex. $1$] The charges are integer multiples of a common
unit $e$, with the mean density $\rho$ not an integer multiple of
$e$.  (This is the most obvious case, with $\gamma = 0$.)
\item[Ex. $2$]  Lattice systems of {\em odd} integer charges with the mean
$\rho$
not equal to an odd integer.  (To cover this case, choose $e=2$ and
 $\gamma = 1$.)
\item[Ex. $3$]
Lattice systems of charges $\pm 1$ (zero not allowed)  which fluctuate,
i.e., the $q_k$ are neither identically $-1$ nor identically $1$.
(This is a special case of $2.$, of relevance to the spin models mentioned
in the introduction.  Note that in this situation there
is no need to add a separate restriction on $\rho$.)

 \end{enumerate}
The last example is intended also to show that there is some room for
maneuver in applying Theorem~\ref{thm:2}.  If one regards the
charge values $\pm 1$  as integers, with $\gamma =0$ and $e=1$, then condition
{\em c.\/} requires that $\rho \neq 0$.  However, if one presents the
two values as $1 + 2 n$  (with $\gamma = 1$ and $e = 2$),
then the restriction on $\rho$ is  $\rho \neq 1 $ (mod 2), but this is
automatically satisfied if the charges fluctuate.

\masubsect{Remarks}

It is natural to explore by how much one can weaken the conditions
which imply translation symmetry breaking.
Following are some remarks concerning such questions (1-4), and other
matters.

\noindent{\emph 1)\/} The condition that $\rho \ne 0$, in the general part
of Theorem~\ref{thm:1}, and the condition  {\em c.\/}, eq. (\ref{eq:alpha}),
in Theorem~\ref{thm:2}, are needed there.  A relevant example is the
one-dimensional two-component Coulomb system with charges $\pm e$,
of equal  densities.  This system, in any of its $\Theta$ states, has
bounded charge fluctuations but it is also mixing under
translations~\cite{Lenard,AM,af} and thus has no cyclic factor (the
correlations of all bounded local functions decay there exponentially
rapidly).  A discrete version of this example is obtained by partitioning
this system into lattice cells.  The state of this lattice model is mixing,
as in the continuum model. The model satisfies conditions {\em a.\/} and
{\em b.\/}, but not {\em c.\/} --- since $\rho = \gamma = 0$ for this model.

\noindent{\emph 2)\/} The discreteness of the matter or the charge
configurations is crucial ---
without this condition it is trivial to construct measures on field
configurations $X_{\omega}(x) $
in $d=1$ with bounded fluctuations and arbitrarily good
mixing behavior.  To do so, one may start from a random field
with  strong clustering properties whose configurations
$u_{\omega}(x)$ are differentiable functions, and take
$X_{\omega}(x) := du_{\omega}/dx$.

\noindent{\emph 3)\/}  The boundedness of the fluctuations cannot be
relaxed much beyond the tightness condition.
For the spin chain with $S=1/2$, i.e., the Heisenberg
antiferromagnet, there is no translation symmetry breaking even though the
growth of the variance of the block spins is subnormal,
increasing only logarithmically with the block length.  The same
appears to be true
for a system of point particles on a line with logarithmic
interactions, and it can be verified by explicit calculation
at special values of $\beta$, e.g. $\beta =2$ --
for which the points correspond to the eigenvalues of
a random Hermitian matrix  sampled from the
Gaussian ensemble and suitably scaled (the Dyson
distribution)~\cite{Mehta}.

\noindent{\emph 4)\/} We do not see straightforward extensions of our
results to higher dimension.  It would be natural to guess that the proper
extension of the boundedness of the variance in particle number, in one
dimension, might be that the variance is of the order of the surface area.
In fact, J. Beck~\cite{Beck} proved that the variance in the particle
number in a ball of radius $r$, averaged over $r$ uniformly distributed in
an interval $(0,R)$, must grow at least like $R^{d-1}$, and such a rate is
realized in some cases of the OCP.  However, examples of the OCP in $d \ge
2$ dimensions show that surface-rate fluctuations are compatible with good
mixing properties.  More explicitly, the OCP in $d \geq 2$ dimensions,
placed in a cube of side $L$ with periodic boundary conditions, will, for
small reciprocal temperatures $\beta$ and small densities $\rho$, have a
limit as $L \to \infty$ which is translation and rotation invariant with
good decay of correlations \cite{Imbrie,BM} and with the particle numbers
having variance of the order of the surface.  (The OCP correlations can in
fact be computed explicitly for $d=2$ at $\beta= 2$ and arbitrary densities
$\rho$: the truncated correlations of $n$ particles are then found to decay
like $\exp[-\alpha D^2]$, where $D$ is the distance between two subsets of
the particle configuration, maximized over the partitions into two
parts~\cite{Choquard,Janco}.)

\noindent{\emph 5)\/} While in the introduction we focused on non-trivial
examples of translation invariant particle measures in $d=1$ with bounded
fluctuations, one should note that there are also many trivial ones.  Just
take the points of the integer lattice and rigidly shift them randomly over
a unit interval with uniform weight.  Or, after shifting, place each
particle with uniform density inside an interval of unit length centered on
the shifted lattice points. In the first example the periodic structure is
clearly visible: the configurations are periodic.  It is less obvious, but
still not difficult to see, that also in the second example there is a
nontrivial periodic structure, though in this case the particle
configurations are not themselves periodic.  (We remark, as an aside, that
the variance on the left side of (\ref{eq:variance}) is minimized, for each
I, by the shifted integer lattice.)

\noindent{\emph 6)\/} In view of the interest in the ``Gibbs state''
condition~\cite{EntFerSok,MS}, let us note that it follows from our results
that the measures satisfying the conditions of our
Theorems~\ref{thm:1} and \ref{thm:2} do not
admit representation as Gibbs states with potentials
for which the interaction across a boundary  is bounded.  The reason
is that in one dimension this boundedness
condition implies uniqueness of the Gibbs measure, which excludes the
possibility of cyclic decomposition~\cite{Gg}.

\bigskip

\bigskip

\masect{ Derivation}
\label{sect:derivation}

The proof of the results in Section~\ref{sect:statement}
rests on two separate observations.
The first is that, under the tightness condition, the configuration of the
charges, centered  by the subtraction of
 its asymptotic mean $\rho$, is the distributional derivative of a process,
 ${\mathcal E}_{\omega}(x)$, which is a covariant functional of the charge
 configuration.  In the second step we make essential use of the atomic
 character of the point-charge process, and in particular of the charge
 constraints, to conclude the existence of a cyclic factor, with cyclic
 function $\phi(\omega)$ of Definition~\ref{def} given by a suitable
 fractional part of ${\mathcal E}_{\omega}(0)={\mathcal E}(\omega)$, namely the
 function $\phi(\omega)$ with values in $[0,2\pi )$ such that
\begin{equation}
\phi(\omega) \ = \ 2 \pi {\mathcal E}(\omega)/ e \quad
  (\mbox{mod $2\pi$})\; .
 \label{eq:theta}
 \end{equation}

The terms used above are defined as follows.
A process ${\mathcal E}_{\omega}(x)$ is said to be a {\em covariant}
functional of $\omega$ (in the sense of behaving covariantly under
translations)
if it satisfies:
\begin{equation}
{\mathcal E}_{\omega}(x) \ = \ {\mathcal E}_{T_x \omega}(0) \; ,
\end{equation}
(equivalently:
$ {\mathcal E}_{\omega}(x+a) \ = \ {\mathcal E}_{T_x\omega}(a)  \; $),
in which case it is determined by the function
\begin{equation}
{\mathcal E}(\omega) \ \equiv \ {\mathcal E}_{\omega}(0) \;  .
\end{equation}
This stationary process is the {\em antiderivative} (or, the
``primitive'') of the charge distribution,
``centered'' in the sense described above, if
\begin{equation}
Q_\omega \left( (a,b] \right)-\rho (b-a) \ = \
{\mathcal E}_{\omega}(b) - {\mathcal E}_{\omega}(a)\; .
\label{eq:defE}
\end{equation}

More generally, let $\omega$ now be any stationary random locally finite
signed measure on $\R$, and write $F_{\omega}(a,b)$ for $\omega((a,b])$. Let
$X_{\omega}(x)$ be its (possibly generalized) ``charge'' density field,
defined through its
(possibly formal) integrals over intervals $\int_a^b X_\omega(x) \,
dx \ = \ F_{\omega}(a,b)$. We shall refer to such an $X_\omega(x)$
as a stationary {\it locally} (weakly) {\it integrable\/} process on $\R$.
We say that a
covariant functional ${\mathcal E}_{\omega}(x)$ is the antiderivative of
$X_\omega(x)$, and that $X_\omega(x)$ is the derivative of ${\mathcal
E}_{\omega}(x)$, if
\begin{equation}
F_{\omega}(a,b) \ = \  {\mathcal E}_{\omega}(b) - {\mathcal E}_{\omega}(a)\; .
\label{eq:defder}
\end{equation}

Note that if a stationary locally integrable process $X_\omega(x)$ is the
derivative of a covariant functional ${\mathcal E}_{\omega}(x)$, then its
integrals over intervals $F_{\omega}(a,b),\  -\infty < a < b < \infty $, form
a tight family of random variables. Of key importance to us is
that the converse is also true.
This fact ---~as we learned from Y. Peres after informing him of our
derivation~--- is related to an established
result in ergodic theory, expressed there as a relation between
tight cocycles and coboundaries (see below).
In this part of the argument the discreteness of the charges does not play
any role, and the statement which we need can be formulated as
follows.

\begin{thm} A  stationary  locally integrable process $X_{\omega}(x)$,
on $\R$, is the derivative of a stationary process ${\mathcal
E}_{\omega}(x)$ given by a covariant functional
if and only if its integrals over intervals
\begin{equation}
 F_{\omega}(a,b)  \ = \  \int_a^b  X_\omega(x) \, dx \
\label{eq:Q}
\end{equation}
with $ - \infty < a < b < \infty$, form a tight family of random variables.
\label{thm:derivative}
\end{thm}

Let us note that it follows from
the theorem that if a process $X_{\omega}(x)$ has any stationary
antiderivative $Y_{\tilde\omega}(x)$ at all---possibly defined only on an
extension of the original probability space and thus not determined by
$X_{\omega}$---it must also have a stationary covariant antiderivative
${\mathcal E}_{\omega}(x)$, since the integrals of $X_{\omega}$ over
intervals then obviously form a tight family.  It is also worth
stressing that the stationary antiderivative of $X_{\omega}(x)$,
whose existence is guaranteed by Theorem \ref{thm:derivative} when the
tightness condition is satisfied, is completely determined, in a
translation invariant manner, by the $X$-process alone.

For one-dimensional Coulomb systems, the antiderivative of the charge
configuration (centered by the subtraction of the asymptotic mean $\rho$)
is, up to an overall shift by a constant, the electric field.
While its meaning is clear  for finite systems, its existence as a
well-defined, and covariant, function of the charge configuration
is not immediately  obvious  for an infinite system~\cite{AM}.

To be more explicit, let us add that if the variables
$\{ F_{\omega}(a,b) \}$ are not only tight but also $L^1$--bounded, satisfying
\begin{equation}
\E(|F(a,b)|) \le K
\label{eq:uniform}
\end{equation}
with $K < \infty$
independent of $a$ and $b$, then
the stationary antiderivative can be chosen as:
\begin{equation}
{\mathcal E}_{\omega}(x) \ = \ E -\lim_{t\to \infty}  \int_0^t (1-\frac st)
X_\omega(s+x)\,ds \; ,
\label{eq:explicit}
\end{equation}
where the integral is guaranteed to converge almost surely through an
application of the ergodic theorem to the process ${\mathcal
E}_{\omega}(x)$ (see the first remark after the proof
Theorem~\ref{thm:derivative}; see also \cite{AM}), and the constant $E$ is
the mean value, $ \E({\mathcal E}_{\omega}(x) ) \ = \ E$.  (If $ {\mathcal
E}_{\omega}(x) $ is regarded, as it is in this paper, merely as giving a
covariant antiderivative of $X_{\omega}(x)$, then the value of the constant
$E$ in \eq{eq:explicit} is of course arbitrary.  However, this
term is essential in any situation where
${\mathcal E}_{\omega}(x)$ has significance in its own right, as it does
for Coulumb systems.  In particular, it plays an important role in the
analysis of the $\Theta$-states mentioned in Section 4.b.)

\masubsect{Existence of stationary anti-derivatives}

A natural context for Theorem~\ref{thm:derivative} is the theory of
cocycles, which has been developed within ergodic theory.
For a measure preserving transformation $T: \Omega \rightarrow
\Omega$,  a cocycle
is a sequence of functions $F_{0}(\omega), F_{1}(\omega),\ldots$
($\omega \in \Omega$) of the form
\begin{equation}
F_{n}(\omega) \ = \  F_{0}(\omega) \ + \
\sum_{j=0}^{n-1} f(T^j \omega ) \; .
\end{equation}
The cocycle is coboundary if $f(\cdot)$ is of the form:
\begin{equation}
f(\omega) \ = \ g(T \omega) - g(\omega) \; .
\end{equation}
In the latter case the cocycle is tight,
as a sequence of random variables, since then
\begin{equation}
F_{n}(\omega)-F_{0}(\omega) \ = \  g(T^n \omega) - g(\omega) \; .
\label{eq:coboundary}
\end{equation}
The converse, derived by Schmidt~\cite{Sch}, is also true:

\begin{thm} (\cite{Sch})
Any {\em tight} cocycle is a coboundary.
\label{thm:cocycle}
\end{thm}

As mentioned above, we owe the reference to Y. Peres.
The $L^2$ version of the result is a yet older result of
Leonov~\cite{Leo}; a recent generalization is found in
ref.~\cite{Aa-W}.

Theorem~\ref{thm:derivative} is a continuum analog of
Theorem~\ref{thm:cocycle}, from which it can be easily derived.  However,
for the completeness of presentation we shall sketch a direct proof.

\begin{proof_of}{Theorem\ref{thm:derivative}}
Given that the argument is not that different from the derivation of
Theorem~\ref{thm:cocycle}, we permit ourselves to present here just a brief
summary.

We start by introducing a convenient extension of the space $\Omega$ over
which the random process $X_{\omega}(x)$ is defined.  Let $\widetilde
\Omega = \Omega \times \R $ and let $\tilde\mu(d\widetilde \omega) $ be a
probability measure on $\widetilde \Omega$ having $\Omega$-marginal $\mu$ .
Writing the points in $ \widetilde \Omega $ as $\widetilde \omega =
(\omega,\, Y)$, we extend the translations $T_x$, originally defined on
$\Omega$, into a flow on $ \widetilde \Omega $ by:
\begin{equation}
\widetilde T_x (\omega,\, Y) \ =
\ \left( T_x \omega,\, Y +F_{\omega}(0,x)\, \right) \; .
\label{eq:t-tilde}
\end{equation}
Notice that on the enlarged space the covariant functional defined by
\begin{equation}
Y_{\widetilde \omega}(x) \ \equiv \  Y(\widetilde T_x \widetilde
\omega )  \; ,
\end{equation}
with $Y(\widetilde  \omega )$  the second coordinate of
$\widetilde \omega$,
provides an antiderivative of $X_{\omega}(x)$,
since
\begin{equation}
Y_{\widetilde \omega}(x) - Y_{\widetilde \omega}(0) \ =
\ F_{\omega}(0,x) \; .
\end{equation}
Moreover, if $\tilde\mu(d\widetilde \omega) $ is stationary under the flow
$\widetilde T_x$, then the random field
$Y_{\widetilde \omega}(x)$, on the probability space $\{\widetilde
\Omega,\tilde\mu\}$, is stationary under $\widetilde T_x$.  This field,
however, need not yet provide us with the covariant functional ${\mathcal
E}_{\omega}(x)$ of Theorem\ref{thm:derivative}, since it
may depend not only of $\omega$ but also  on the entire $\widetilde
\omega$.

The proof of Theorem~\ref{thm:derivative} is in two steps.  First, it is
established that there indeed exists a probability measure $\tilde
\mu(d\widetilde \omega) $ on $\widetilde \Omega $ such that: {\em i.\/} the
$\Omega $ marginal of $\tilde\mu$ is
$\mu(d \omega)$, and {\em ii.\/} $\tilde\mu$ is
stationary under  $\widetilde T_x$.  Then, using the stationarity
---~which permits the application of the ergodic theorem~--- we construct a
modified field on $\widetilde \Omega$ which depends only on the first
coordinate, $\omega$.  It is this field which yields the desired ${\mathcal
E}_{\omega}(x) $.

To prove the first claim, we let $\nu_L (d\widetilde \omega) $ be the
probability measure for which
$\omega$ has the distribution $\mu(d\omega)$ and the conditional
distribution of  $Y$, given $\omega$, is that of the random
variable $Y =   F_\omega (-u,0)  $
with $u$ sampled over $[0, L]$ with the uniform probability
distribution ( $d u /L$ ).
It is easy to see that $\nu_{L}(\cdot) =\frac1L\int_0^L
\widetilde T_u \nu_0\;du$, where $\nu_0=\mu\times\delta_0$,  is close to
being invariant under $\widetilde T_x$, with the variational distance between
$\widetilde T_x \nu_{L}(\cdot) $ and $\widetilde \nu_{L}(\cdot) $
bounded by $2x / L$.

It is also easy to see that the distribution of $Y$ under the
measures $\nu_L(\cdot)$ is tight:
\begin{eqnarray}
\P_{\nu_L} (\,|Y|\ge t\,)  \
& = & \ \frac1L\int_{0}^{L} \P_{\mu}(\, |F_{\omega }(-u,0)| \ge t\, ) \;du
  \nonumber \\
& & \nonumber
\\ &\ \le & \
\sup_{u>0} \P_{\mu}(\, |F_{\omega }(-u,0)| \ge t\, )\; .
\label{eq:tight}
\end{eqnarray}
Under the tightness assumption of Theorem \ref{thm:derivative},
the right hand side vanishes as $t\to \infty$.

The above considerations suggest that compactness and continuity arguments
can be invoked to prove that the sequence of probability measures
$\nu_{L}(\cdot)$ has a convergent subsequence, as $L \to \infty$, and that
the limiting measure is strictly stationary under the flow $\widetilde
T$. The existence of a suitable limiting measure can indeed be proven using
a topology which is natural for $\omega$ in the present set-up (it is at
this point that we leave the argument at the level of a sketch), for which
\eq{eq:tight} implies that the sequence of measures $\nu_{L}(\cdot)$ is
tight, and thus has a convergent subsequence, with limit $\tilde\mu$.

Clearly the $\Omega-$marginal of $\tilde\mu$ is $\mu$. That $\tilde\mu$ is
also stationary would be immediate were $\widetilde T_x$ continuous on
$\widetilde \Omega$. However, there is here a slight complication in that
the second component on the right-hand-side of \eq{eq:t-tilde} is not
continuous in $\omega$.  Nevertheless, the set of configurations $\omega$
for which $\widetilde T_x \widetilde \omega$ is discontinuous---those
configurations having atoms at $0$ or at $x$---has $\mu$-measure $0$, (an
easy consequence of the stationarity of $\mu$).  Using this fact, the
conclusion of stationarity follows.

The above implies that in the larger space
$\widetilde \Omega$, the process
$X_{\omega}(x)$ has a stationary antiderivative $Y_{\widetilde
\omega}(x)$.  While the antiderivative
thus obtained need not be determined by
$\omega$ alone, its stationarity  implies that it has the necessary
asymptotic statistical regularity  to yield
an antiderivative of the field $X_\omega$ which is a  covariant
function of $\omega$ alone.  This can be done
by selecting from among the one-parameter family of possible
antiderivatives (differing only by an overall shift)
the one whose median value, averaged along the
positive $x$-axis, is set at $Y=0$.

The median level for $Y(\cdot)$ is defined by
\begin{equation}
M[Y] \ = \inf_{M} \left\{  M \, :
\lim_{L\to \infty} \frac1L \int_0^L
{\mathcal I}[Y(x) \ge M] \, dx  \ \le 1/2
\,\right\} \; ,
\end{equation}
provided the limit exists.  (The symbol ${\mathcal I}[\ \cdot\ ]$
represents the indicator function.)
For $Y_{\widetilde \omega}(\cdot)$, which is sampled with a stationary
distribution, the existence of the limit (simultaneously for a countable
collection of $M$) follows by the ergodic theorem.  Using the median, we define
\begin{equation}
{\mathcal E}_{\widetilde \omega}(x) \ =
\ Y_{\widetilde \omega}(x) - M[Y_{\widetilde \omega}]  \; .
\end{equation}
Because $M[Y+\mbox{Const.}]=M[Y]+\mbox{Const.}$, ${\mathcal E}_{\widetilde
\omega}$
depends only on $\omega$ (and not on the other coordinate of  $ \widetilde
\omega $), and thus it defines a suitable covariant antiderivative of
$X_{\omega}(x)$.
\end{proof_of}

\noindent{\bf Remarks:}

\noindent 1. Under the additional assumption that \eq{eq:uniform} holds,
the stationary antiderivative $Y_{\widetilde\omega}$ constructed in the
first step of the argument above satisfies
$\E[Y_{\widetilde\omega}(0)]<\infty$, as may easily be seen from the
construction of $\tilde\mu$.  One may then alternatively define an
antiderivative functional ${\mathcal E}$ using the average $\bar Y$ of $Y$,
$\bar Y=\lim_{L\to \infty} \frac1L \int_0^L Y(x) \, dx $, in place of the
median $M[Y]$. Moreover, for this antiderivative we may obtain an explicit
formula: Averaging the expression ${\mathcal E}_{\omega}(0) - {\mathcal
E}_{\omega}(u)= -F_\omega(0,u)$ with respect to $du/L$ over the interval
$(0,L]$ and taking the limit $L \to \infty$, using the fact that, by
construction, $\lim_{L\to \infty} \frac1L \int_0^L {\mathcal E}_\omega(u)
\, du =0$, one arrives, in fact, at the formula given by \eq{eq:explicit}
(see ref.~\cite{AM} for further discussion of \eq{eq:explicit}).

\smallskip
\noindent 2.  In the case of the spin models discussed in the
introduction (example {\bf b}), the antiderivative $Y_{\omega}(x)$
corresponds to the total `charge' to the left of $x$ in those clusters
which are split by $x$.

\smallskip
\noindent 3. Suppose $X_{\omega}(x)$ is ergodic. Then it is easy to
see---either by using the covariant antiderivative of Theorem
\ref{thm:derivative} or directly---that a stationary antiderivative
$Y_{\tilde\omega}(x)$ of $X_{\omega}(x)$ is a covariant functional of
$\omega$ alone if and only if $Y_{\tilde\omega}(x)$ is ergodic. Moreover,
two ergodic antiderivatives of $X_{\omega}(x)$ differ by an absolute
constant. Thus if $X_{\omega}(x)$ is ergodic, instead of proceeding to step
2 as described above, we could obtain a covariant antiderivative ${\mathcal
E}$ by simply decomposing $\tilde\mu$, corresponding to the antiderivative
$Y_{\tilde\omega}(x)$ from step 1, into its ergodic components and choosing
any one of these. Moreover, using the ergodic theorem, it follows from the
manner of construction of $\nu_L$ that the process $Y_{\tilde\omega}(x)$
from step 1 is a mixture of the antiderivatives ${\mathcal E}(x) - y$ with
weights $\mu_{{\mathcal E}(0)}(dy)$, i.e., with $y$ random, with distribution
that of ${\mathcal E}(0)$. In particular, the distribution of
$Y_{\tilde\omega}(0)$ is that of the difference of two independent copies
of ${\mathcal E}(0)$. Note also that it follows that all limits of $\nu_L,\
L\to\infty,$ must agree, independent of which convergent subsequence is
chosen, so that in fact $\lim_{L\to\infty}\mu_L$ itself exists,
without passage to a subsequence.

\bigskip

\masubsect{Proof of the main results}

\begin{proof_of}{Theorem~\ref{thm:1}}
It suffices to consider the point-charge processes, since any point process
may be regarded as a special case with unit charges.  Under the assumption
of tight fluctuations, Theorem~\ref{thm:derivative} says that there is a
measurable function ${\mathcal E}(\omega)$ of the charge configuration such
that
\begin{equation}
{\mathcal E}(T_x\omega) \ = {\mathcal E}(\omega) + Q_{\omega}((0,x]) - \rho
x \; .
\end{equation}
Since the charge $Q_{\omega}((0,x])$ assumes only values which are integer
 multiples of $e$, we may conclude that $\phi(\omega)\equiv
 \frac{2\pi}{e}{\mathcal E}(\omega)\ ({\rm mod}\  2\pi)$ defines a cyclic
 factor as described in Theorem~\ref{thm:1} and Definition \ref{def}.
\end{proof_of}

\begin{proof_of}{Theorem~\ref{thm:2}}
The argument in the discrete case is similar to the above, with the only
notable difference being that the charge $Q_{\omega}(\, (0,x]\, )$ now includes
$\gamma x$ in addition to an integer multiple of $e$. Thus in this case the
``antiderivative functional'' ${\mathcal E}(\omega)=g(T\omega)$ provided by
Theorem~\ref{thm:cocycle} satisfies
\begin{equation}
{\mathcal E}(T_x\omega) \ = {\mathcal E}(\omega) + (\gamma- \rho) x \quad ({\rm
mod}\ e)
\end{equation}
and hence  $\phi(\omega)\equiv
 \frac{2\pi}{e}{\mathcal E}(\omega)\ ({\rm mod}\  2\pi)$ defines a cyclic
 factor as described in Theorem~\ref{thm:2} and Definition \ref{def}.
\end{proof_of}

\masect{Extensions}
\label{sect:newapp}

\masubsect{Jellium tubes}

Our discussion of one-dimensional systems applies also to elongated tubes
which are locally of higher dimension, but are of finite cross section and
infinite in one direction.  An example of such a system is the three
dimensional OCP with the flux lines of the three dimensional field
restricted to stay in the tube.  (this is an idealized situation, possibly
mimicking some high contrast dielectric materials, or a Kalutza-Klein model
with only one unconfined dimension.)  If the variances of the total charge
in tubes of length $L$ stays bounded, as may be expected under the Coulomb
interactions, then the theory presented here implies translation symmetry
breaking, along the unconfined direction.  In such a situation the
projections of the positions of the point charges along the free direction
yield a one dimensional system of the type discussed in this paper. For
$d=2$ this system is considered in \cite{Choquard} for the explicitly
solvable case $\beta=2$.

\masubsect{$\Theta$-states of one-dimensional Coulomb gas}

Unlike the OCP the one-dimensional Coulomb gas consisting of a continuum
system of point charges of value $\pm e$, with respective densities
$\rho_+=\rho_-$, does not break translation symmetry, see Remark 1) of
Section 2.c.  This system does however exhibit another form of
non-uniqueness of the Gibbs state: it admits a one-parameter family of
infinite-volume translation invariant Gibbs states, indexed by the
fractional part $(\Theta)$ of the boundary charge ``imposed at infinity''
\cite{AM,af}.  While at first glance this ``anomaly'' appears to be of
a different kind than what is discussed here, let us point out that it can
also be viewed as a remnant of the translation symmetry breaking in
asymmetric charge systems.

The charge-symmetric model may be arrived at as a limit of Coulomb systems
with a uniform background of small charge density $\rho_{bg}=- \epsilon$,
and $\rho_+ = \rho_- +\epsilon $.  In this situation our results do apply, and
prove that the system exhibits translation symmetry breaking, the
corresponding infinite volume Gibbs states having a cyclic factor of period
$\lambda=1 / \epsilon$.  For each $\epsilon$ the resulting cyclic
components $\mu_\theta^{(\epsilon)}$ (of equation (\ref{eq:cc}) ) are
periodic under translations, with period $1/ \epsilon$.  Moreover, the
dependence of $\mu_\theta^{(\epsilon)}$ on $\theta$ can be so chosen that
when the asymmetry parameter $\epsilon$ is taken down to zero, these
converge (locally) to a one-parameter family of translation invariant
states (i.e., probability measures on the space of configurations) which
form the $\Theta$-states of the symmetric Coulomb gas.  The proof can be
obtained using the methods of ref.~\cite{AM}.



\noindent {\large \bf Acknowledgments\/} \\
We thank Yuval Peres for calling our attention to the relation of
our work with the theory of tight cocycles [24-26].
MA gratefully acknowledges the hospitality accorded him at
Rutgers University.   The work was supported in part by the NSF grants
PHY-9971149, DMR-9813268 and AFOSR Grant F49620--98--1--0207.

 \addcontentsline{toc}{section}{References}

\end{document}